# Multimodal fusion for sea level anomaly forecasting

Guosong Wang[12], Xidong Wang[1], Xinrong Wu[2], Kexiu Liu[2], Yiquan Qi[1],Chunjian Sun[2],
Hongli Fu[2]

1 College of Oceanography Hohai University, Nanjing 210098, China

2 National Marine Data and Information Service, Tianjin 300171, China

**Abstract**: The accumulated remote sensing data of altimeters and scatterometers have provided a new opportunity to forecast the ocean states and improve the knowledge in ocean/atmosphere exchanges. Few previous studies have focused on sea level anomaly (SLA) multi-step forecasting by multivariate deep learning for different modalities. For this paper, a novel multimodal fusion approach named MMFnet is used for SLA multi-step forecasting in South China Sea (SCS). First, a grid forecasting network is trained by an improved Convolutional Long Short-Term Memory (ConvLSTM) network on daily multiple remote sensing data from 1993 to 2016. Then, an in-situ forecasting network is trained by an improved LSTM network, which is decomposed by the ensemble empirical mode decomposition (EEMD-LSTM), on real-time, in-situ and remote sensing data. Finally, the two single-modal networks are fused by an ocean data assimilation scheme. During the test period from 2017 to 2019, the average RMSE of the MMFnet (single-modal ConvLSTM) is 4.03 cm (4.51 cm), the 15th-day anomaly correlation coefficient is 0.78 (0.67), the performance of MMFnet is much higher than those of current state-of-the-art dynamical (HYCOM) and statistical (ConvLSTM, Persistence and daily Climatology) forecasting systems. Sensitivity experiments analysis indicates that, compared with a set of based single models, the MMFnet, which added CCMP SCAT products and OISST for SLA forecasting, has improved the forecast range over a week and can effectively produce 15-day SLA forecasting with reasonable accuracies. Take wind speed and SST anomaly as additional input, the MMFnet has better forecasting ability for large-radius eddies in the open sea and coastal regions, overcoming weakness of single forecasting models. In an extension of the validation over the North Pacific Ocean, MMFnet can calculate the forecasting results in a few minutes, and we find good agreement in amplitude and distribution of SLA variability between MMFnet and other classical operational model products.

# 1 Introduction

The South China Sea (SCS) is the sea where typhoons (Wang et al.,2014), mesoscale eddies (Tuo et al.,2019), internal waves and other weather and marine phenomena occur frequently, which puts forward higher requirements for environmental and operational ocean forecasting. Sea level anomaly (SLA) forecasting was acknowledged to provide ocean thermal structure to support accurate anti-submarine warfare (ASW) acoustic prediction performance (Burnett et al.,2014). With the increasing importance of marine battlefields, the rapid and intelligent SLA forecasting with limited computing resources is becoming increasingly important in oceanography (Li et al.,2020). In 1992, ERS-1 and T/P satellite remote sensing first observed the global SLA distribution, representing the first possible eddy-resolution global ocean forecasting observation system (Smedstad et al.,2003). In recent years, various kinds of remote sensing, in-situ observations, model data, and other earth system data have proliferated. However, the SLA forecasting has not improved rapidly with the increase in available data. The SLA forecasting technology is mainly focuses on numerical forecasting and empirical statistical forecasting, the most widely used global operational ocean models include HYCOM (The HYbrid Coordinate Ocean Model) (Chassignet et al.,2009) and NEMO (The Nucleus for European Modelling of the Ocean) (Madec,2015). Currently, The National Real-Time Ocean Forecasting System (RTOFS) of the National Center for Environmental Prediction (NCEP) is based on 1/12 ° HYCOM model of eddy-resolution(Mehra et al.,2010), and use the 3D multivariate data assimilation method NCODA (Cummings,2005) to obtain the 8-day forecasting Marine prediction. The NEMO model also has been widely applied in the marine forecasting, For example, the French government's Mercator-Ocean nowcast systems use Oceanic General Circulation Models (OGCM) to predict sea level up to 10 days ahead (Drevillon et al.,2008). With the development of high-performance computing and observation systems, more scientific challenges have been brought to the study of numerical model physical processes, parameterization schemes, and data assimilation algorithms (Bauer et al.,2015).

As the result of the massive accumulation of multivariate observation data and the growth of computing power, the extensive application of advanced deep learning (Goodfellow et al.,2016; LeCun et al.,2015; Schmidhuber,2015) has provided new methods and ideas for SLA forecasting, The latest study shows that the marine and meteorological forecasting is more accurate and energy-efficient in the parameterization of key physical processes (Jiang et al.,2018; Bolton et al.,2019; Gentine et al.,2018). Currently, recurrent neural networks (RNNs) (Hochreiter et al.,1997) and convolutional neural networks (CNNs) (Lecun et al.,1998) have achieved state-of-the-art results on a number of future time series forecasting benchmarks. Yoo-Geun Ham et al. used both the CMIP5 output and the reanalysis data in transfer learning to train a CNN first on historical simulations, advance the skillful forecast of ENSO events up to 1.5 years (Ham et al.,2019). At the same time, The CNN model predicts the detailed zonal distribution details of sea surface temperatures well, overcoming a weakness of dynamical forecast models. Qin Zhang et al. made a 7-day forecast of 5 random points in the Bohai Sea based on the high-resolution NOAA SST data of the first 30 days (Zhang *et al.*, 2017). Caixia Shao use the Holt-Winters and ARIMA models to fit the interannual and residual terms of SSHA in SCS and the valid forecast time of SCS SSHA is about 7 months.(Shao et al.,2015). The RNN encoder-decoder model based on attention mechanism also reached a new level (Bahdanau et al.,2014).

On the other hand, deep learning for spatiotemporal sequence forecasting is essential for a wide range of scientific studies and real-life applications like precipitation nowcasting of the Radar Echo Extrapolation (REE) problem. In 2015, Xingjian Shi et al. proposed a convolutional long-short-term memory network (ConvLSTM) structure for spatiotemporal sequences, which converts the step-by-step prediction problem into a spatiotemporal sequence forecast problem under an end-to-end learning framework(Shi et al.,2015). The multi-layer ConvLSTM network was trained for the first time on Doppler radar echo data. Compared with the traditional optical flow method, the accuracy of precipitation nowcasting in Hong Kong has been significantly improved. This ConvLSTM model has become a seminal work in this area. Xingjian Shi et al. proposed an upgraded trajectory GRU (TrajGRU) network in 2017, which actively learns the characteristics of local changes in recursive connections, and solves the problem of constant space-time structure caused by the use of convolution loop connections in ConvLSTM networks(Shi et al.,2017). Subsequently, Chunyong Ma et al. constructed a SLA forecasting network based on ConvLSTMs that choosing the general learning strategies for Iterated Multi-step (IMS) estimation, which learns a one-step ahead forecaster and iteratively applies it to generate multi-step predictions, to preserve the sharpness of the predicted frames. The results show that the forecast error of the ConvLSTM model on the seventh day is 3.28 cm, and the matching rate for eddies with a diameter greater than 100 km is about 60% (Ma et al.,2019). Admittedly, the stacked ConvLSTM architecture is proved powerful for supervised spatiotemporal learning, the memory cells that belong to the layers are mutually independent and updated merely in time domain. Recent advances in ConvRNNs include the introduction of external memory, PredRNN(Wang et al.,2017) and PredRNN++(Wang et al.,2018), which models spatial and temporal representations in a unified memory cell and convey the memory both vertically across layers and horizontally over states. This is different to the simple ConvLSTM model, which just can recall and update the temporal correlation. And they are shown to be better than TrajGRU in both of the REE and video frame prediction tasks (Wang et al.,2019).

Due to the complex spatial and temporal relationships within the multivariate and the potential forecasting limit, there are three major challenges of SLA multi-step forecasting by multivariate deep learning for different modalities. The first challenge is how to learn a model for multi-step forecasting. The IMS approach is easy to train and less computationally expensive, while the Direct Multi-step (DMS) approach, which is directly optimizes the multi-step forecasting, can avoid the error drifting problem.

Choosing between DMS and IMS involves a trade-off among forecasting bias, estimation variance, the length of the prediction horizon and the model's nonlinearity (Taieb et al.,2014). The second challenge is how to adequately model the spatial and temporal structures between different modalities. With the development of computer vision and natural language processing, Multimodal Fusion has gradually become a valuable research issue in academia and industry (Jin et al.,2017; Li et al.,2017). Multimodal fusion integrates information from two or more modalities (such as spatial, time series, etc.) with more comprehensive information characteristics to obtain robustness and consistent results. Few previous studies have utilizing deep learning to gain the shared representative information between different modalities to resolve the problem of SLA forecasting. The main reason is that it is difficult to learn good representations for both short-term topography dependency and long-term multivariate relations. The third challenge is how to application of multivariate. Regional sea level variability, both temporal and spatial, is dominated by locally ocean and atmosphere variability (Miles et al.,2014), thus extreme SLA forecast in SCS require accurate knowledge of regional multiple variables variability. Unfortunately, to the best of our knowledge, the state-of-the-art deep learning-based model seldom application of multivariate for the SLA forecasting, despite of its importance.

In this paper, we explore a novel Multivariate Multimodal Fusion network (MMFnet) that integrates an improved LSTM network, which is decomposed by EEMD (EEMD-LSTM) into ConvLSTM, which applying DMS approach (ConvLSTM) by an ocean data assimilation scheme. The MMFnet, which is included wind and SST as additional input variables, can effectively and directly produce 15-day SLA forecasting with reasonable accuracies.

## 2 Data and Methodology

### 2.1 Data

In this paper, three long-term daily satellite remote sensing products are used for training, test, and verification in deep learning. The multi-satellite altimeter SLA dataset is distributed by the Copernicus Marine Environment Monitoring Service (CMEMS), it provides a consistent and homogeneous catalogue of products for near real time applications; The gridded wind vector analysis data is from the Cross-Calibrated Multi-Platform (CCMP) V2.0 data set(Atlas et al.,2011; Atlas et al.,1996),provide a consistent, gap-free long-term time-series of ocean surface wind vector analysis fields; The SST is from the NOAA Optimum Interpolation (OI) Sea Surface Temperature (SST) V2 High Resolution Dataset(Reynolds et al.,2010). All the remote sensing data is from January 1, 1993 to December 31, 2018 and is on a 1/4 deg global grid. The 80×80 gridded data of SLA, SST anomaly (SSTA) and wind speed anomaly (SPDA) in the SCS (105°E - 125°E, 5°N -25°N) were extracted, respectively

For data preprocessing, we first zero-center each feature of the data, and then normalize its range of values to the range [-1, 1]. After normalization and standardization, the speed of gradient descent can be accelerated. It can use a larger learning rate to perform gradient propagation more stably and improve the generalization ability of the neural networks. It is worth noting that the preprocessing operations in this paper can only be performed on the training set data, and the algorithm is applied to the validation set and the test set after training. Then in order to obtain the disjoint subset required for training, test, and verification, the data set in this paper is divided into the consecutive sequence from 1993 to 2013 as the training set, the consecutive sequence from 2014 to 2015 as the verification set, and the consecutive sequence from 2016 to 2017 as the test set. We slice the sequence with a 30-days sliding window. Each sequence consists of 30 gridded data, with SLA, SSTA and SPDA as input in the first 15 days and SLA as output in the last 15 days. Then, the total 12,222 sequences are split into a training set of 8,148 samples, a validation set of 730 samples and a test set of 731 samples.

## 2.1 Methodology

In this paper, a novel multimodal fusion approach named MMFnet is used to produce 15-day SLA forecast, propose to gain the shared representative information between different modalities of SLA in SCS. The framework of MMFnet consists of three parts:

First, an improved multivariate ConvLSTM network applying DMS approach (ConvLSTM-DMS), which is trained on daily remote sensing observations from T-14 to T, is used for the gridded SLA forecasting in SCS. The ConvLSTM-DMS network is trained by a 4-layer ConvLSTM network which applying DMS approach with $3 \times 3$ convolution kernel and 128, 128, 64 and 32 hidden states respectively. The filter of the first convolution layer is used to detect low-order features such as eddy edges, angles, curves, etc. With the increase of convolutional layers, the characteristics of eddy motion detected by corresponding filters become more complicated, the network has multiple stacked ConvLSTM architecture, so SLA forecasting under complex dynamic environments is feasible.

Secondly, an improved multivariate LSTM network, which is decomposed by the ensemble empirical mode decomposition (EEMD-LSTM) and trained on in-situ daily remote sensing observations on remote sensing data from T-14 to T, is used for the in-situ SLA forecasting in the sensitive areas. The areas where the local large values of the root-mean-square error (RMSE) are located represent the sensitive areas in SCS. We selected 20 number of observations in the shelf area and the Luzon Strait where the errors are most concentrated and expect them to have a considerable impact on forecast skills. The ensemble empirical mode decomposition (EEMD) algorithm decomposes each of time series sequence signal into 5 intrinsic mode functions (IMFs) at different time scales to obtain more realistic and physically meaningful signals and reduce the impact of noise (Huang et al.,2019; Liu et al.,2019). We employ the technique to examine certain aspects of this nonlinear and nonstationary time series. Finally, the output of each IMF component are synthesized for in-situ SLA forecasting. The EEMD-LSTM in this paper contains 4-layer LSTM network, and have been used to solve many real-life sequence modeling problems (Sagheer et al.,2019; Chao et al.,2018).

Finally, We explore a novel MMFnet network that integrates in-situ EEMD-LSTM forecasting network into the ConvLSTM-DMS network by an ocean data assimilation scheme, which uses an inverse distance weight method of the SLA background gradient information to learn the spatiotemporal relationship information of different modes(Hongli et al.,2013), 15-day SLA forecasting experiments are conducted in SCS. Spatiotemporal sequence fusion forecast is a key issue in multimodal. The contributions of temporal and spatial models are different, simply combining may cause the temporal-mode prediction results to be masked by spatial model. The ocean data assimilation scheme effectively prevents the excessive unphysical projection of observational information, and thus improves assimilation quality greatly. Full details of the ocean data assimilation scheme are provided in Hongli Fu et al. (2013).

We train all deep learning networks by minimizing the cross-entropy loss of time backpropagation (BPTT) and RMSProp with a learning rate of $10^{-3}$ and a decay rate of 0.9. The optimizer is Nadam. The use of L2 regularization makes the network more inclined to use all input features instead of relying on some small features in the input features to control the neural network overfitting. The small batch gradient descent method is used to update the parameters in batches, which reduces the randomness and saves the calculation. Also, we perform early-stopping and directly optimizes the 15-step SLA forecasting. Unless otherwise specified, the batch size of each iteration is set to 32 and the parameters are adjusted separately to obtain the best performance. All experiments are implemented in TensorFlow and conducted on NVIDIA Quadro P6000 GPUs with 50G Memory.

## 3 Experiments and Evaluation

In this paper, root-mean-square error (RMSE) and anomaly correlation coefficients (ACC) were used to evaluate the performance of the network.

$$RMSE(f,x) = \sqrt{\frac{1}{N}\sum_{i=1}^{N}(f-x)^2} \qquad (1)$$

$$ACC(f,x) = \frac{\sum_{i=1}^{N}(f-\overline{f})(x-\overline{x})}{\sqrt{\sum_{i=1}^{N}(f-\overline{f})^2}\sqrt{\sum_{i=1}^{N}(x-\overline{x})^2}} \qquad (2)$$

Where, $N$ is the total number of statistical objects, $f$ is the forecasted SLA, and $x$ is the satellite altimeter SLA dataset. $\overline{f}$ is the average forecasted SLA, $\overline{x}$ is the average satellite altimeter SLA . The value of ACC ranges between 0 and 1, and the larger value of the ACC or the smaller value of the RMSE, the greater the improvement of the SLA forecast skills.

### 3.1 Assessment of multimodal fusion network

In this section, we first compare the ConvLSTM-DMS and EEMD-LSTM on the test data set to gain some basic understanding of the behavior of single-model deep learning network. As a strong competitor, we also included the widely used univariate ConvLSTM network, which is follow the IMS structure, as baseline model (ConvLSTM-IMS).

Notably, SCS has both a wide continental shelf, deep-sea basins and steep continental slopes. Except for the shallow water in the north and west coastal regions, the water depth in the middle open ocean regions is above 2000m. The topography changes is a common problem that had a considerable impact on the forecast ability of the network (Masina et al., 1994), it is vital to validate model results in both coastal regions and in open ocean regions.

Figure 3 shows the average RMSE verification of SLA on daily time scales between the altimeter data and ConvLSTM networks (both IMS and DMS) and MMFnet network on the test set during 2016-2017, clearly presenting that the ConvLSTM-DMS and ConvLSTM-IMS networks both success has the ability to learn the temporal and spatial relationships associated with time series data in the central SCS. The large RMSE values are mainly distributed in the coastal regions of northern SCS within the 10-day forecast, which may be related to the relatively high dynamic height caused by the accumulation of seawater on the west coast caused by coastal currents and Ekman transport. More than 10-day forecast, the RMSE grows more rapidly beyond that time in Luzon Strait, where is characterized as having high mesoscale eddy activity (Wang et al.,2012).

The RMSE decreases to 5.10 from 6.17 cm after applying the DMS approach in ConvLSTM-DMS. Thus, ConvLSTM-DMS has general improvement in representing of the SLA in both coastal regions and in open ocean regions of SCS. However, it is still cannot improve the forecast skills of SLA in the sensitive areas. These is mainly because it is difficult to do feature extracting in different region modalities by ConvLSTM network without time series analysis methods, which are prone to cause modal deviations, leading to a decline in forecasting capabilities. We conjecture that the use of a fixed-kernel is a bottleneck in improving the performance of this basic ConvLSTM architecture. If the states are viewed as the hidden representations of moving eddies, then the ConvLSTM with a larger kernel should be able to capture faster motions while one with a smaller kernel can capture slower motions (Xingjian et al.,2015). It is almost impossible for the single-model deep learning networks like ConvLSTM to make accurate and comprehensive forecast of SLA in different seas at the same time.

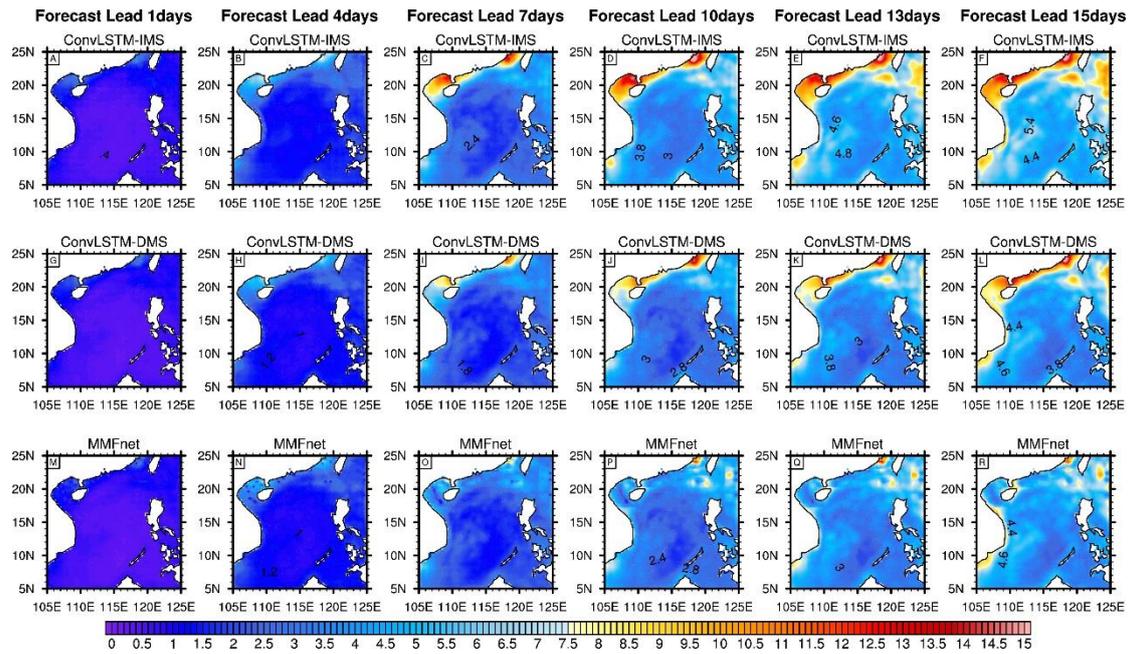

Fig. 1 The average RMSE at each grid point on the test set from 2016 to 2017. (A–F) The RMSE maps from the first day to the 15th day of ConvLSTM-IMS, (G–L) The RMSE maps from the first day to the 15th day of ConvLSTM-DMS. (G–L) The RMSE maps from the first day of MMFnet. The output forecast are shown at three days intervals.

We observe that there is a strong non-linearity and non-stationary in the shelf shallower than 100 m and the Luzon Strait (Fig. 2), which has a strong impact on the performance of the network. Over the shelf area in SCS, however, the SLA data still contains aliases from tides and internal waves (Yuan et al.,2006). Since the error distribution of SLA is in line with the area where the sea level changes dramatically in the SCS region, it is necessary to develop an in-situ forecasting network methodologies that help to decrease the RMSE in shelf area and Luzon Strait.

The average RMSE on the test set of each in-situ forecast in Table 1 shows that, the improved EEMD-LSTM in-situ forecasting network is significantly superior to ConvLSTM-DMS, reducing the average RMSE score from 6.97cm to 1.75cm. An example of the high daily SLA variability is observed at the west coast of Taiwan Strait. At these locations, the RMSE is 9.09 cm during 2016-2017 when comparing daily SLA time series from altimeter data. However, the RMSE decreases to 2.67 from 9.09 cm after applying the improved EEMD-LSTM in-situ forecasting network to the time series. A notable decrease in RMSE values is evident that EEMD-LSTM is able to forecast SLA with an acceptable accuracy at nearly all locations, especially in west coast of Taiwan Strait and Luzon Strait.

Meanwhile, These additional in-situ forecast result had been assimilated by a data assimilation system to form a more reliable forecast state, The RMSE distribution of MMFnet show that (Fig. 3h-l), it is beneficial to fused the two single-model deep learning networks, The statistics are also substantially improved by using the MMFnet network to eliminating the errors in coast of northern SCS and Luzon Strait at the same time, which has been supplied to the model for a more accurate forecast.

Table 1 Geographical details (Longitude, Latitude) and the regional average RMSE of 20 observations in the coastal and the Luzon Strait on the test set from 2016 to 2017. (units: cm)

| Sensitive areas | abbreviation | Longitude | Latitude | ConvLSTM-DMS | EEMD-LSTM |
|---|---|---|---|---|---|
| West coast of Taiwan Strait | S1 | 119.50 | 24.00 | 7.95 | 2.17 |
| | S2 | 120.00 | 24.50 | 7.72 | 2.18 |
| | S3 | 118.38 | 23.88 | 10.94 | 2.84 |
| | S4 | 118.88 | 24.62 | 9.76 | 3.47 |
| Coast of Guangdong province | S5 | 116.50 | 22.84 | 6.90 | 1.44 |
| | S6 | 114.50 | 22.00 | 5.41 | 1.42 |
| | S7 | 111.80 | 21.00 | 6.27 | 1.44 |
| | S8 | 112.50 | 21.50 | 7.44 | 2.40 |
| Gulf of Beibu | S9 | 108.12 | 20.38 | 7.28 | 2.02 |
| | S10 | 108.75 | 21.50 | 7.42 | 2.17 |
| | S11 | 107.12 | 20.38 | 6.60 | 1.80 |
| | S12 | 107.20 | 19.50 | 6.00 | 1.50 |
| | S13 | 107.50 | 18.50 | 5.48 | 1.15 |
| | S14 | 108.20 | 18.00 | 5.32 | 1.10 |
| | S15 | 109.50 | 20.50 | 8.15 | 2.16 |
| East of Luzon Strait | S16 | 124.62 | 19.62 | 5.02 | 1.05 |
| | S17 | 124.62 | 21.50 | 5.10 | 0.98 |
| | S18 | 124.88 | 22.12 | 5.97 | 0.95 |
| West of Luzon Strait | S19 | 119.85 | 21.85 | 5.07 | 1.05 |
| | S20 | 118.63 | 21.12 | 9.59 | 1.64 |

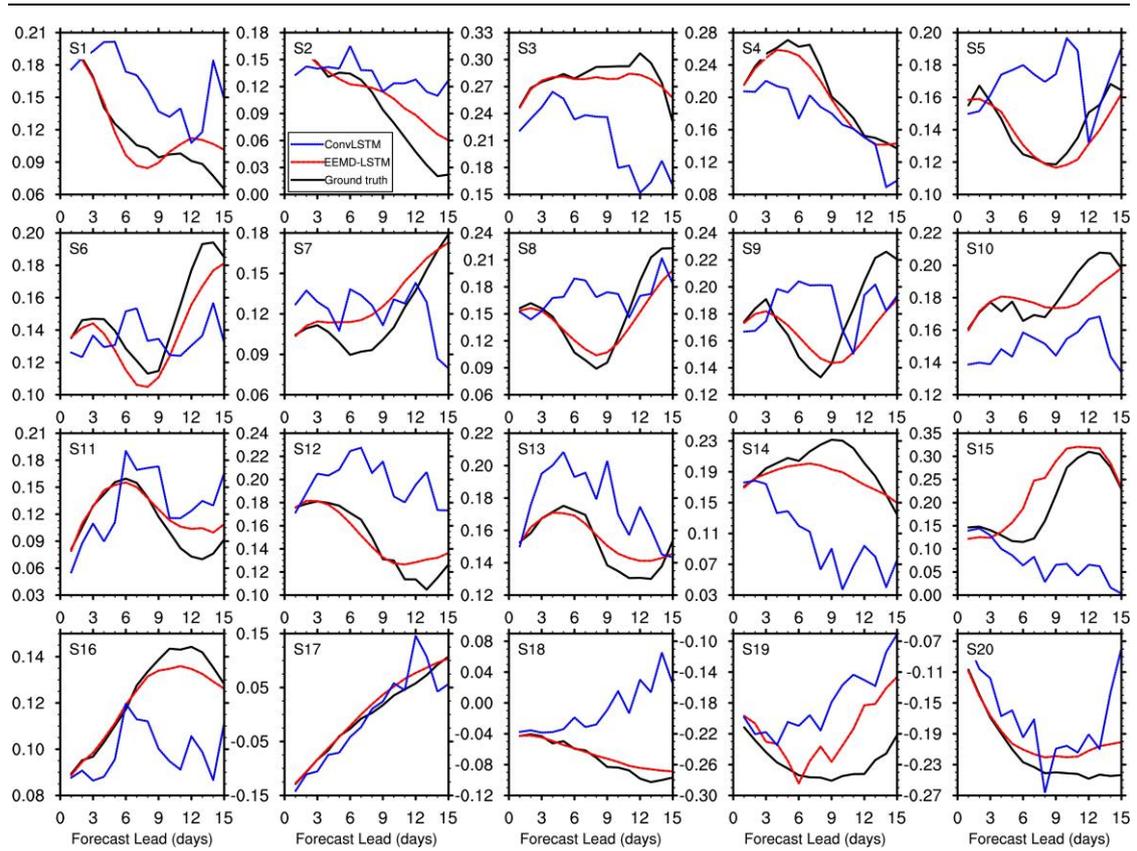

Fig. 2 A 15-day forecast example of ConvLSTM-DMS and EEMD-LSTM against the remote sensing validation dataset. (units: m)

## 3.2 Sensitivity analysis of multivariate

The application of multiple input variables for the SLA forecasting loading a challenge. In this subsection, we further investigate the impact of the multiple input variables on the multimodal fusion for SLA forecasting. We conducted a set of sensitivity tests for the MMFnet models. Two satellite remote sensing data, SST anomaly (SSTA) and 10-m wind speeds (SPDA,UA10 and VA10) , were included as additional input variables in the tests, We train three 4-layer MMFnet models with 1(SLA), 2(SLA,SSTA) and 3(SLA,SSTA, SPDA) input variables respectively. It is worth noting that, the deep neural network uses the same model structure and uses the best parameters for SLA forecasting; therefore, to impartially compare the forecast uncertainties of SLA in different experiments.

As seen from the RMSE and ACC of sensitivity tests in 2016-2017 (Table 2),

MMFnet with multivariate (MMFnet1, MMFnet2 and MMFnet3) is significantly superior to the baseline model ConvLSTM-IMS on the test sets with remote sensing altimeter over the SCS. MMFnet2 with only 2 input reduces the the regional average RMSE to 2.794 cm on the test set. By included SPDA and SSTA as additional input variables, we further decline the sequence RMSE from 2.918 cm down to 2.784 cm. Over time, the ACC gradually decreases, and the average ACC of MMFnet3 gradually decreases from 0.997 on the first day to 0.831 on the 15th day. The median ACC is 0.86 and the median RMSE is 0.89. MMFnet2 is slightly worse than MMFnet3, but there is not much difference between the two. It is noted that MMFnet with multivariate (MMFnet2 and MMFnet3) usually results in RMSE values lower than MMFnet1 for all years. Although the multivariate may not be able to resolve the physical process, the more input variables involved, the greater the improvement of the SLA forecast skills obtained.

Table 2 Sensitivity tests of MMFnet and baseline model ConvLSTM-IMS. We report the regional average RMSE (units: cm) and ACC of generated sequences averaged across the test sets. Lower RMSE or higher ACC denotes better forecast accuracy.

| Model | Input Variables | forecast day 1 | | forecast day 7 | | forecast day 15 | | average | |
|---|---|---|---|---|---|---|---|---|---|
| | | RMSE | ACC | RMSE | ACC | RMSE | ACC | RMSE | ACC |
| MMFnet1 | SLA | 0.671 | 0.997 | 2.722 | 0.946 | 4.853 | 0.808 | 2.918 | 0.920 |
| MMFnet2 | SLA,SSTA | 0.670 | 0.997 | 2.611 | 0.948 | 4.627 | 0.830 | 2.794 | 0.926 |
| MMFnet3 | SLA,SSTA, SPDA | **0.670** | **0.997** | **2.604** | **0.948** | **4.618** | **0.831** | **2.784** | **0.927** |
| ConvLST M-IMS | SLA | 0.791 | 0.995 | 3.443 | 0.919 | 6.169 | 0.771 | 3.618 | 0.898 |

We use the RMSE as metrics to evaluate the forecast results and the corresponding step-by-step quantitative comparisons are presented in Fig. 3. The spatial and monthly mean RMSE results show that, The ConvLSTM-IMS model seemed to not work well in more than one week (Fig. 4a), while the MMFnet models in both tests remain stable over time, only with a slow and reasonable decline. There is no significant difference between the MMFnet with multivariate, this is not surprising considering that these MMFnet networks with multivariate are strongly affected by convolution kernel size or hidden states layers number that are same or fixed in the network structure. However, compared to MMFnet1, the most obvious effect of the multivariate is a change in October. While the regional average RMSE of MMFnet1 is about 4cm at forecast day 5, and 6 cm at forecast day 10 in October 2017(Fig. 4b), the forecast lead-time has improved at forecast day 8,(8) and 12(13) after applying the multivariate to MMFnet2 (MMFnet3) (Fig. 4c-d).

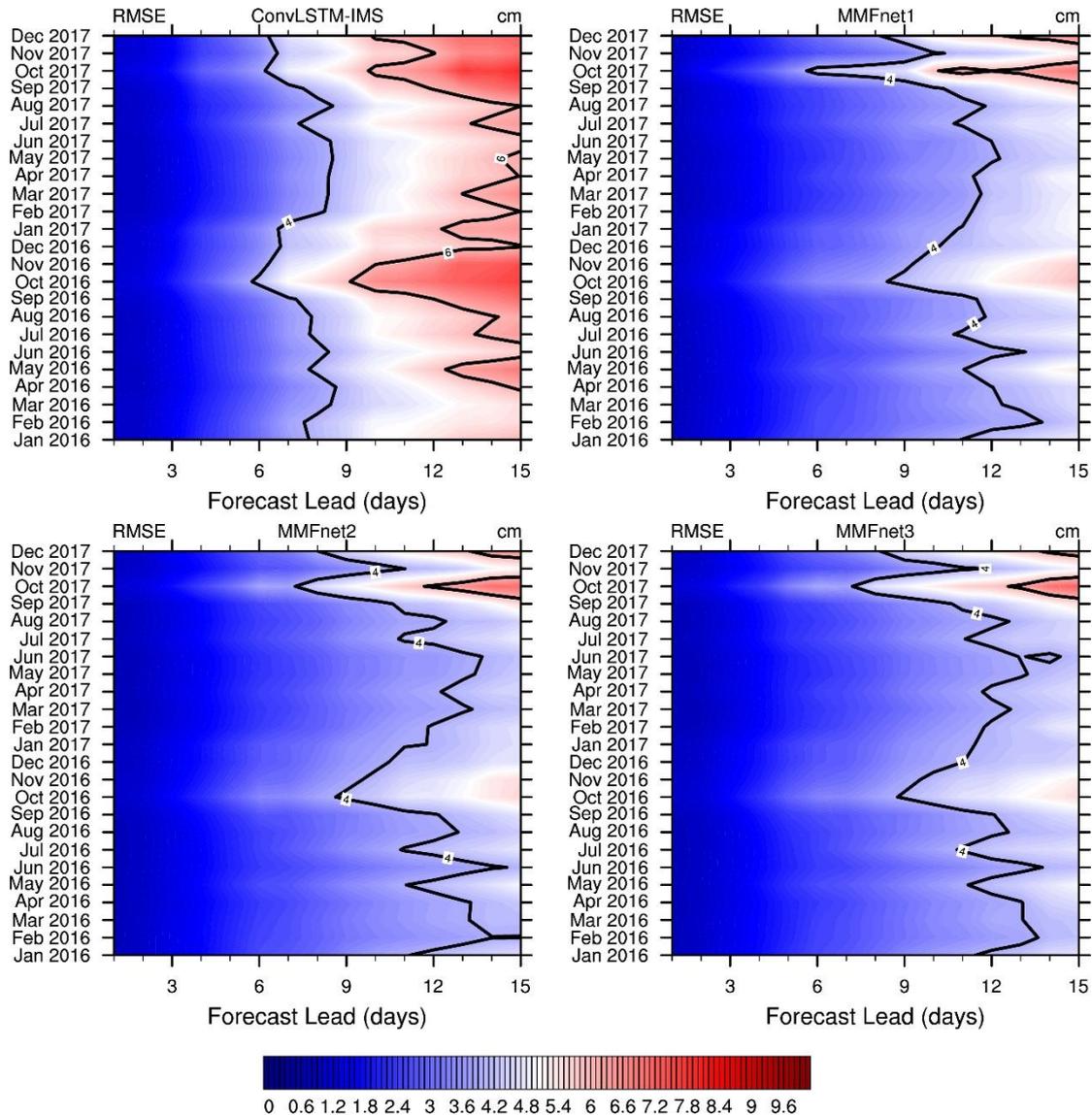

Fig. 4 Step-wise of monthly mean RMSE comparisons of MMFnet with multivariate (MMFnet1, MMFnet2 and MMFnet3) and baseline model ConvLSTM-IMS on the test sets with altimeter observation over the SCS. The plots show the monthly mean RMSE over 731 15-day forecasts for the period 1 January 2016 to 31 December 2017. The solid curve is 4.0-contour and 6.0-contour, respectively. (units: cm)

Another example of the sensitivity analysis of multivariate is shown the difference between MMFnet2 and MMFnet1, MMFnet3 and MMFnet1 to illustrate the spatial variation of the effect of the multivariate (Fig. 5). The ACC of MMFnet1 with single input variable does not differ significantly for the first three days. However, forecast lead-time for more than one week, the quality measurements in the forecast is more sensitive to the number of input variables involved in the networks. The MMFnet2 and MMFnet3 networks are significantly reduces forecast errors and improves the forecast range in central SCS and East Luzon Strait. In part, this is due to the MMFnet1 can only solve a rough and smooth pattern of SLA for lack of marine and meteorological forcing. It is difficult to maintain the SLA structure and strength more than one week. In contrast, the MMFnet with 2 and 3 input variables separates atmospheric and ocean factors into different feature layers, enabling MMFnet to determine the role of atmospheric and ocean factors to produces a stable SLA forecasting. As shown in Figure 6, it is interesting to see that MMFnet3 had quite similar performance with MMFnet2 for the first 12 steps, While the overall relative performance of the MMFnet2 compared to MMFnet3 is mixed, the MMFnet3 forecasts are more accurate in both the central SCS, East Luzon Strait and Taiwan Strait, but less in coastal regions, especially at the longer forecast times.

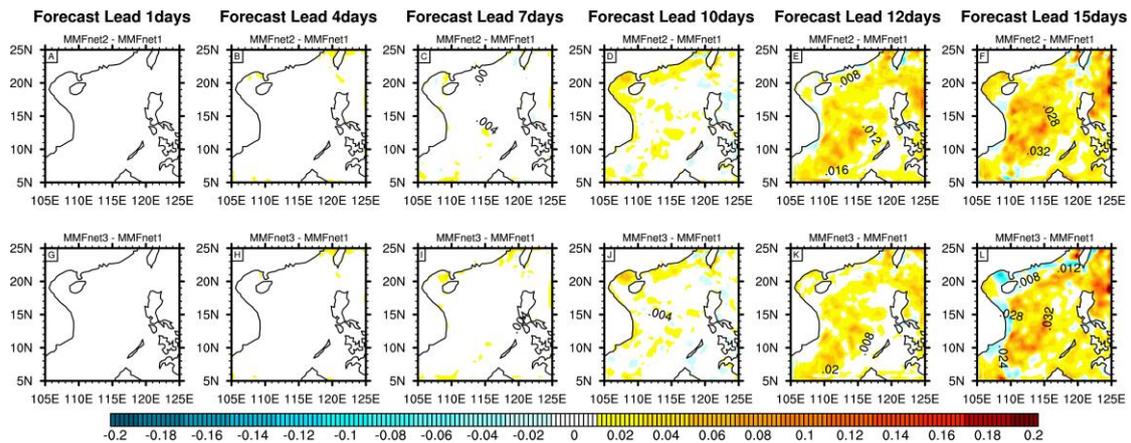

Fig. 5 The RMSE maps from the first day to the 15th day on the test set from 2016 to 2017 (units: cm). (a) Daily RMSE difference between MMFnet2 and MMFnet1, and (b) daily RMSE difference between MMFnet3 and MMFnet1.

## 3.3 Comparisons of SLA forecast among various model products

Using the same test set and metrics, We investigated further into the performance of MMEnet with HYCOM model product and three sets of benchmark forecasts, including two statistical forecasts (Persistence and daily Climatology), and ConvLSTM-IMS . The persistence is based on the assumption that the initial state of the oceanic variables will persist for the entire time of the forecast (Briggs,2007) and they represent an economical forecast system(Shriver et al.,2007). Persistence has a long history of use as a benchmark to decide whether a forecast technique (model) has forecast value (Arcomano et al.,2020).

The HYCOM-based Global Ocean Forecasting System (GOFS) 3.1, which uses the Navy Coupled Ocean Data Assimilation (NCODA) system (Cummings,2005) for data assimilation, generates a nowcast out through a seven-day forecast. The daily real-time SSHA forecast from HYCOM model were compared with remote sensing altimeter, the Mean Dynamic Topography (MDT) used in HYCOM is the temporal mean of the SSH above the Geoid over a period 1993-2012, the resolution is interpolated to a 0.25 ° grid.

We use the RMSE as metrics to calculate the results of the above 5 models on the test set. The corresponding step-by-step quantitative comparisons are presented in Figure 3. Daily climatology has poor forecasting performance, while persistence forecasting and ConvLSTM-IMS have a certain forecasting ability within one week. This is mainly because the production and development process of large diameter eddies is more stable and less affected by non-linear factors. Forecasts within one week have strong spatial correlation, that is, the SLA movement in local areas is highly consistent. However, with the increase of forecast range, the performance of persistence and ConvLSTM-IMS decreases rapidly and nearly the same for the last 5days. Moreover, the complex structure of ConvLSTM is still plagued by the problem of the disappearance of the gradient. Through the back propagation of time, the amplitude of the gradient decays exponentially. The dependence on long-term prediction and training can easily cause the problem of the disappearance of the gradient. Consistent with previous research results (Xu et al., 2011), the seven-day forecast average RMSE of HYCOM reached 14cm across the whole SCS, and this is mainly because the errors in MDT have significant impacts on forecast results. Since HYCOM does not assimilate SSH anomaly field directly, it is diagnosed from the prognostic bottom pressure and internal density fields (Halliwell et al.,2014). The MDT must match that contained in the time mean altimeter data, which is a nontrivial problem (Cummings, 2014).

The MMFnet3 has a forecasting capacity of at least 15 days in SCS. Compared with other models, MMFnet3 turns out to be more accurate for long-term forecast and significantly outperforms all previous dynamic forecasting and statistical methods, has improved the forecast range over one week. This is mainly because the MMFnet3 is trained end-to-end for SLA forecasting, and the spatiotemporal convolution structure of the network successfully learns the spatiotemporal sequence forecast characteristics of SLA. For statistical forecasting methods, it is difficult to find a reasonable way to update future SLA fields and train all data end-to-end yet.

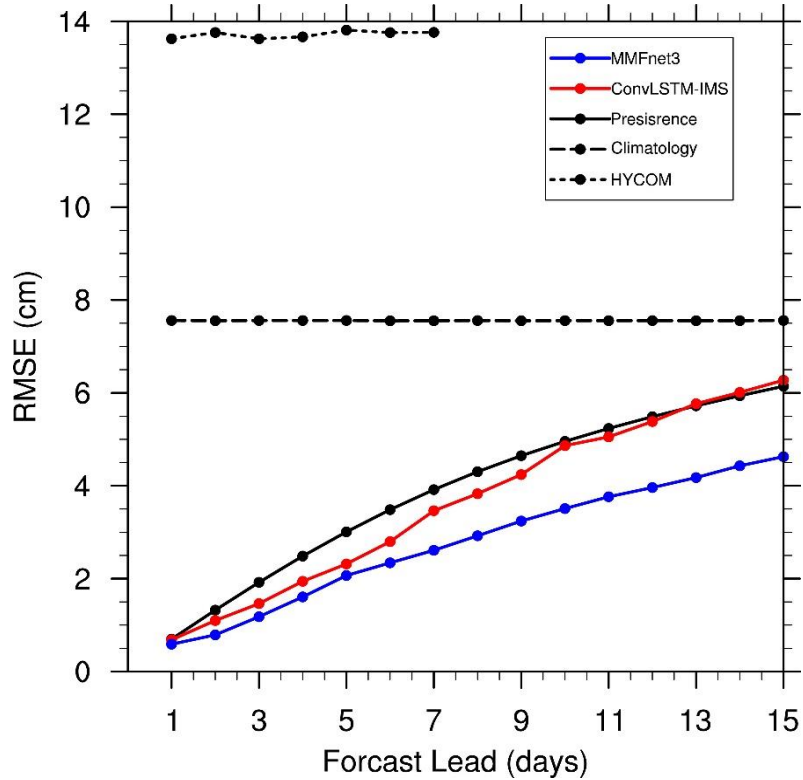

Fig. 6 The regional average RMSE comparisons of sensitivity tests of MMFnet and baseline model ConvLSTM-IMS on the test sets with altimeter observation over the SCS. The plots show the RMSE over 731 15-day forecasts for the period 1 January 2016 to 31 December 2017. The red curve is the ConvLSTM-IIMS; the blue curve is the MMFnet; the black solid, dashed and dotted curve is the forecasts of persistence (i.e., no change from the initial state), daily Climatology (SLA daily mean climatology from 1993-2017) and HYCOM. (units: cm)

Eddy nowcatsing is the most challenging task in the SLA nowcasting, thus the modeling spatial deformation is important. We select a typical SLA sequences with relatively complicated spatiotemporal variations (in both eddy trajectory and propagation) as an example, a qualitative comparison is given in Figure 4. ConvLSTM-IMS is less computationally expensive in training, but prone to focus on spatial appearances and relatively weak in capturing long-term motions due to the accumulation of errors caused by using the prediction result of the previous time step as the input of the next time step in multiple iterations. The eddy nowcatsing of ConvLSTM-IMS in the central and western SCS tend to blurry for more than one week. Although dynamic and statistical forecast can produce clearer results than deep learning networks, they produce more false forecast and are often less accurate than deep learning networks.

It can be seen intuitively that the RMSE increases slightly with the prolongation of the forecast period in MMFnet, and the blurry is significantly reduced in the forecast for more than one week. It not only significantly reduces the error of the northern coast of SCS, but also clearly forecast the propagation and evolution of eddies in the western SCS and the eastern Luzon Strait. This is mainly due to two reasons. One reason is that, MMFnet3 network can handle boundary conditions. There are a large number of eddy motion samples in the training set in Luzon Strait, the MMFnet3 , which it is beneficial to fused the two single-model deep learning networks, can learn the SLA spatiotemporal characteristics of different regions with different network during training, and memorizes detailed appearances, as well as long-term motions, thus make reasonable forecast in the boundary. However, it is difficult to forecast the spatiotemporal complexity in ConvLSTM-IMS. Another is that, the MMFnet3 network adds marine and meteorological forcing as inputs for end-to-end training, the SLA structure and intensity forecast have been improved, especially for the last 7 days. Overall the comparisons of SLA variability, the MMFnet3 show good agreement in amplitude and distribution and outperforms all baseline models and shows superior forecasting power both spatially and temporally.

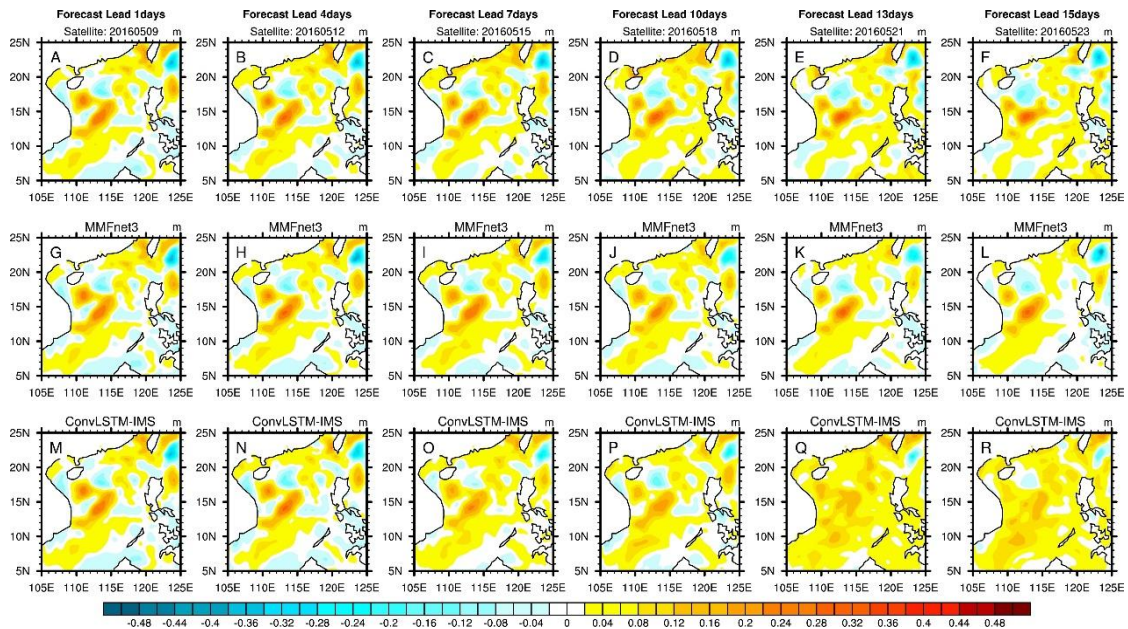

Fig. 7 A forecast example on the test set. (A-F) altimeter observation ;( G-I) MMFnet3 and (M-R) ConvLSTM-IMS. 15-day SLA forecasting by observing 15 previous real time remote sensing products. The output forecast are shown at three days intervals. (units: m)

## 4 Summary and Conclusion

In this paper, a novel multimodal fusion approach named MMFnet is presented for the problem of SLA forecasting by multivariate deep learning for different modalities in SCS. Results presented in this paper reveal that The MMFnet model, which fuses two single-model deep learning networks (ConvLSTM-DMS and EEMD-LSTM), can effectively produce 15-day SLA forecasting with reasonable accuracies, as evidenced by 731 15-day forecasts performed during 2016-2017.The approach of assimilating SLA of in-situ EEMD-LSTM forecasting network via the ocean data assimilation scheme is shown to greatly reducing the RMSE in both coastal regions and in Luzon Strait regions, improve skill in SLA forecasting. A set of sensitivity analysis of multivariate results indicated that, with an appropriate selection of input variables, MMFnet model, which added CCMP SCAT products and OISST for SLA forecasting, has improved the forecast range over one week and reduces the obstacles of October forecasting than univariate ConvLSTM network .Overall, we find indication that MMFnet turns out to be more accurate for long-term forecast and superior than those of current state-of-the-art dynamical forecast (HYCOM) and statistical forecast (ConvLSTM-IMS, Persistence and daily Climatology). At the same time, The MMFnet not only clearly show the ability to forecast the propagation and evolution of eddies in the western SCS and the eastern Luzon Strait, but also overcome the weakness of single-mode network forecasting blurry.

Under the condition of maintaining prediction performance and limited GPU memory training, we propose a 2-layer MMFnet network suitable for $3 \times 3$ convolution kernels for SLA forecasting in the North Pacific. The SLA 7-day forecast results show that based on the trained parameters, MMFnet quickly completes the calculation within 1 minutes, and the forecast results are in good agreement with the SLA intensity and distribution of other numerical model product forecasts. It should be noted that MMFnet is a feasibility and energy efficiency computer code that can run on a variety of computing platforms.

Multimodal fusion deep learning can improve the prediction ability of seasonal prediction and long-distance spatial connection modeling across multiple time scales, solve the problem of fast, intelligent, and accurate prediction in marine battlefield scenarios, and has a wide range of civil and military applications . For example, through the MODAS three-dimensional inversion method and deep learning such as CNN, the relationship between sea surface information and the state of the subsurface layer can be deeply explored, and the information of sea surface temperature and sea surface state of the sea surface layer can be mapped into the underwater dynamic environment to reconstruct the subsurface layer. The three-dimensional structure field provides real-time continuous three-dimensional state information for global marine products. At the same time, the comprehensive application of multi-modal fusion networks, numerical ocean forecasting and physical oceanography method can achieve empowerment and value enhancement, and promote the deep cross-fusion of marine science, high-performance computing and artificial intelligence.